\documentclass[12pt]{article}
\usepackage{latexsym,amsfonts,amssymb}
\makeatletter
\@addtoreset{equation}{subsection}
\makeatother

\begin{document}
\begin{center}
{\Large \bf Notes on Quantum Field Theory in Curved Spacetime:\\
Problems Relating to the Concept of Particles and Hamiltonian
Formalism}
\\[1.5cm]
 {\bf Vladimir S.~MASHKEVICH}\footnote {E-mail:
  Vladimir.Mashkevich100@qc.cuny.edu}
\\[1.4cm] {\it Physics Department
 \\ Queens College\\ The City University of New York\\
 65-30 Kissena Boulevard\\ Flushing\\ New York
 11367-1519} \\[1.4cm] \vskip 1cm

{\large \bf Abstract}
\end{center}

The aim of these notes is to elucidate some aspects of quantum
field theory in curved spacetime, especially those relating to the
notion of particles. A selection of issues relevant to
wave-particle duality is given. The case of a generic curved
spacetime is outlined. A Hamiltonian formulation of quantum field
theory in curved spacetime is elaborated for a preferred reference
frame with a separated space metric (a static spacetime and a
reductive synchronous reference frame). Applications: (1) Black
hole. (2) The universe; the cosmological redshift is obtained in
the context of quantum field theory.

\newpage

\section*{Introduction}

One of the essential features of quantum phenomena is the
wave-particle duality. It is explicitly represented both in
quantum mechanics (eigenstates of coordinate and of momentum) and
in quantum field theory in flat spacetime (annihilation/creation
operators and field modes). Furthermore, in both theories, the
notion of the Hamiltonian is substantial. However, the situation
is different in quantum field theory in curved spacetime. As long
as a generic curved spacetime is considered [1-8], the concepts of
particles and of the Hamiltonian are inconsistent. The reason is
that a generic spacetime has no global structure and, therefore,
no preferred field modes and vacuum state. However, it is
difficult to abandon such vital notions as particle and the
Hamiltonian.

In a generic curved spacetime, a Hamiltonian formulation of
quantum field theory is inappropriate [1], so a Lagrangian
formulation is adopted [1-8]. The situation is different if there
exists a preferred reference frame with a separated space metric:
$ds^{2}=g_{00}(dx^{0})^{2}+g_{ij}dx^{i}dx^{j}$, where $(-g_{ij})$
is a Riemannian metric. Then a Hamiltonian formulation may be
implemented. In [9,10], the scalar quantum field has been
constructed in a special case of cosmic spacetime.

There are two general cases with the above metric: a synchronous
reference frame, where $g_{00}=1,\;g_{ij}=g_{ij}(x^{0},(x^{l}))$;
static spacetime, where
$g_{00}=g_{00}((x^{l})),\;g_{ij}=g_{ij}((x^{l}))$. In this paper,
both cases are treated from a uniform point of view.

The  spin--0, 1, and 1/2 fields and the Hamiltonian are
constructed out of annihilation and creation operators. In the
case of free fields, there are no divergences. The construction
fits into the customary pattern for flat spacetime.

Applications to the universe and the Schwarzschild black hole are
given. For the universe (specifically, in the FLRW model) the
cosmological redshift is obtained in the context of quantum field
theory.

Some issues concerning interaction are considered.

\section{Wave-particle duality}

\subsection{Wave-particle duality in quantum mechanics}

One of the most essential features of quantum phenomena is the
wave-particle duality. We quote Bohm [11]:

``\ldots Bohr wanted to make the wave-particle duality the
starting point of the physical interpretation.

\ldots Bohr developed his conception of complementarity from the
wave-particle duality\ldots: There exist complementary
properties---like position and momentum---and the exact
measurement of one precludes the possibility of obtaining
information of the other. Properties are not actualities; they are
only possibilities for the physical system. These developments
formed the basis of the so-called Copenhagen interpretation of
quantum mechanics.''

The simplest quantum mechanical system is a zero-spin particle.
Operators relating to the particle are constructed out of
$\hat{X}$ (coordinate) and $\hat{P}$ (momentum), a state in the
coordinate representation is the wave function $\psi(x)$. A
measurement of the coordinate operator $\hat{X}$ would result in a
state that has as its classical image a particle. On the other
hand, the system in an eigenstate of the momentum operator
$\hat{P}$ behaves like a wave.

\subsection{Wave-particle duality in quantum field theory}

The simplest quantum field system is a scalar field. Operators
relating to the field are constructed out of the operator-valued
distribution $\hat{\varphi}(x),\;x=\{x^{\mu}:\mu=0,1,2,3\}$, a
(pure) state is represented by a vector of the Hilbert space,
$\Psi\in \mathcal{H}$.

We quote Kuhlmann [12]:

``Many of the creators of QFT can be found in one of the two camps
regarding the question whether particles or fields should be given
priority in understanding QFT. While Dirac, the later Heisenberg,
Feynman, and Wheeler opted in favor of particles, Pauli, the early
Heisenberg, Tomonaga, and Schwinger put fields first\ldots''

We have
\begin{equation}\label{1.2.1}
\hat{\varphi}(x)=\hat{\varphi}^{(+)}(x)+\hat{\varphi}^{(-)}(x),\quad
\hat{\varphi}^{(\mp)}=\hat{\varphi}^{(\pm)\dag}
\end{equation}
in the Heisenberg picture
\begin{equation}\label{1.2.2}
\hat{\varphi}^{(+)}=\sum_{m}\frac{1}{\sqrt{2\omega_{m}}}
f_{m}(\vec{x})\mathrm{e}^{-\mathrm{i}\omega_{m}t}\hat{a}_{m}
\end{equation}
where the space mode
\begin{equation}\label{1.2.3}
f_{m}(\vec{x})=\frac{1}{\sqrt{V}}
\mathrm{e}^{\mathrm{i}\vec{p}_{m}\vec{x}}
\end{equation}
and the frequency
\begin{equation}\label{1.2.4}
\omega_{m}=\sqrt{\mu^{2}+p^{2}_{m}}
\end{equation}
The Hamiltonian is
\begin{equation}\label{1.2.5}
\hat{H}=\sum_{m}\omega_{m}\hat{a}_{m}^{\dag}\hat{a}_{m}
\end{equation}
The wave aspect is represented by the modes
$f_{m}(\vec{x})\mathrm{e}^{-\mathrm{i}\omega_{m}t}$, the particle
aspect by the $\hat{a}_{m},\;\hat{a}_{m}^{\dag}$---in the sense of
integrity, which is manifested in the annihilation and creation of
particles. We quote Dirac [13]:

``A fraction of a photon is never observed.''

Locality is represented by the transformation
\begin{equation}\label{1.2.6}
\hat{a}'_{m'}=\gamma_{m'}{}^m\hat{a}_{m}\,,\quad
F^{'m'}=F^{m}(\gamma^{-1})_{m}{}^{m'}
\end{equation}
\begin{equation}\label{1.2.7}
\hat{\varphi}^{(+)}=\sum_{m}F^{m}\hat{a}_{m}=\sum_{m'}F^{'m'}\hat{a}'_{m'}\,,\quad
F^{m}=\frac{1}{\sqrt{2\omega_{m}}}f_{m}\mathrm{e}^{-\mathrm{i}\omega_{m}t}
\end{equation}
\begin{equation}\label{1.2.8}
\gamma^{\dag}\gamma=\gamma\gamma^{\dag}=I
\end{equation}
with a suitable $\gamma$.

As to the role of annihilation/creation operators in quantum
theory, we quote Weinberg [14]:

``\ldots creation and annihilation operators were first
encountered in the canonical quantization of the electromagnetic
field and other fields\ldots They provided a natural formalism for
theories in which massive particles as well as photons can be
produced and destroyed\ldots

However, there is a deeper reason for constructing the Hamiltonian
out of creation and annihilation operators, which goes beyond the
need to quantize any pre-existing field theory like
electrodynamics, and has nothing to do with whether particles can
actually be produced or destroyed. The great advantage of this
formalism is that if we express the Hamiltonian as a sum of
products of creation and annihilation operators, with suitable
non-singular coefficients, then the $S$-matrix will automatically
satisfy a crucial physical requirement, the cluster decomposition
principle, \ldots which says in effect that distant experiments
yield uncorrelated results. Indeed, it is for this reason that
formalism of creation and annihilation operators is widely used in
non-relativistic quantum statistical mechanics, where the number
of particles is typically fixed. In relativistic quantum theories,
the cluster decomposition principle plays a crucial part in making
field theory inevitable.''

And Tung [15]:

``(Connection Between Representations of Lorentz and Poincar\'e
Groups.) The $c$-number wave functions
$u^{\alpha}(\vec{p}\lambda)\mathrm{e}^{\mathrm{i}px}$ in the plane
wave expansion\ldots are the coefficient functions which connect
the set of operators $\{a(\vec{p}\lambda)\}$, transforming as the
irreducible unitary representation $(m,s)$ of the Poincar\'e
group, to the set of field operators $\Psi^{\alpha}(x)$,
transforming as certain finite dimensional non-unitary
representation of the Lorentz group.

To pursue this group theoretical interpretation of the ``plane
wave solution'' of the wave equation a little further, note that
$u^{\alpha}(\vec{p}\lambda)\mathrm{e}^{\mathrm{i}px}$ carries both
the Poincar\'e indices $(\vec{p}\lambda)$ and the Lorentz indices
$(x,\alpha)$.''

\section{Quantum fields in a generic curved spacetime}

\subsection{The problem of the concept of particles}

In the case of quantum fields in curved spacetime, the situation
changes dramatically. We quote Kay [7]:

``The main new feature of quantum field theory in curved spacetime
(present already for linear field theories) is that, in a general
(neither flat, nor stationary) spacetime there will not be any
single preferred state but rather a family of preferred states,
members of which are best regarded as on an equal footing with
one-another. It is this feature which makes the above algebraic
framework particularly suitable, indeed essential to a clear
formulation of the subject. Conceptually, it is this feature which
takes the most getting used to. In particular, one must realize
that\ldots the interpretation of a state as having a particular
``particle-content'' is in general problematic because it can only
be relative to a particular choice of ``vacuum'' state and,
depending on the spacetime of interest, there may be one state or
several states or, frequently, no states at all which deserve the
name ``vacuum'' and even when there are states which deserve this
name, they will often only be defined in some approximate or
asymptotic or transient sense or only on some subregion of the
spacetime.

Concomitantly, one does not expect global observables such as the
``particle number'' or the quantum Hamiltonian of flat-spacetime
free field theory to generalize to a curved spacetime context, and
for this reason local observables play a central role in the
theory. The quantized stress-energy tensor is a particularly
natural and important such local observable and the theory of this
is central to the whole subject.''

And Wald [8]:

``Major issues of principle with regard to the formulation of the
theory arise from the lack of Poincare symmetry, the absence of a
preferred vacuum state, and, in general, the absence of asymptotic
regions in which particle states can be defined\ldots

The particle interpretation/description of quantum field theory in
flat spacetime has been remarkably successful---to the extent that
one might easily get impression from the way the theory is
normally described that, at a fundamental level, quantum field
theory is really a theory of particles. However, the definition of
particles relies on the decomposition of $\phi$ into annihilation
and creation operators\ldots This decomposition, in turn, relies
heavily on the time translation symmetry of Minkowski spacetime,
since the ``annihilation part'' of $\phi$ is its positive
frequency part with respect to time translations. In a curved
spacetime that does not possess a time translation symmetry, it is
far from obvious how a notion of ``particles'' should be
defined.''

\subsection{Divergences}

In the conventional treatment, even in the case of free fields,
there are divergences in the energy-momentum tensor operator, and
the renormalized operator is only defined up to a finite
renormalization ambiguity [7], [8].

\section{Preferred reference frame}

\subsection{Reference frame with separated space metric}

We now turn to a Hamiltonian formulation of quantum field theory
in curved spacetime. It is based first of all on the selection of
a reference frame with a preferred time coordinate, so that
general covariance is broken. A relevant spacetime manifold is the
product manifold:
\begin{equation}\label{3.1.1}
M^{\mathrm{spacetime}}=T^{\mathrm{time}}\times
S^{\mathrm{space}}\qquad M\ni p=(t,s)\quad t\in T,\;s\in S
\end{equation}
and metric is of the form
\begin{equation}\label{3.1.2}
g=g^{\mathrm{time}}dt\otimes dt+g^{\mathrm{space}}\qquad
g^{\mathrm{space}}=-h
\end{equation}
or
\begin{equation}\label{3.1.3}
ds^{2}=g_{00}(dx^{0})^{2}+g_{ij}dx^{i}dx^{j}\qquad
g_{ij}=-h_{ij}\quad i,j=1,2,3
\end{equation}
where $h=(h_{ij})$ is a Riemannian metric on $S$. Thus, a selected
reference frame is time-orthogonal [16], or that with a separated
space metric.

Note that the choice of spatial coordinates is in general
immaterial.

\subsection{Static spacetime}

The first case when metric is of the standard form (3.1.2) is that
of a static spacetime, where
\begin{equation}\label{3.2.1}
g^{\mathrm{time}}=g^{\mathrm{time}}(s)\qquad h=h(s)\quad s\in S
\end{equation}
or
\begin{equation}\label{3.2.2}
g_{00}=g_{00}((x^{l}))\qquad h_{ij}=h_{ij}((x^{l}))
\end{equation}

\subsection{Synchronous reference frame}

The second case with metric of the standard form (3.1.2) is that
of a synchronous reference frame, where
\begin{equation}\label{3.3.1}
g^{\mathrm{time}}=1\qquad h=h_{t}(s)
\end{equation}
or
\begin{equation}\label{3.3.2}
g_{00}=1\qquad h_{ij}=h_{ij}((x^{\mu}))\quad \mu=0,1,2,3
\end{equation}

\subsection{Preferred reference frame}

The selection of a preferred reference frame amounts to the choice
of a time coordinate. In the case of a static spacetime, the
choice is unique. But in the case of a synchronous reference
frame, the choice is by no means uniquely defined. In that case,
the uniqueness is achieved by the condition that the time $t$ be
that of simultaneous quantum state reduction, i.e., quantum jumps
over all the space $S$. This condition defines a reductive
reference frame.

Thus, a preferred reference frame is either related to a static
spacetime or is a reductive reference frame.

\subsection{Energy-momentum tensor and Hamiltonian}

In semiclassical gravity, the energy-momentum tensor is defined as
\begin{equation}\label{3.5.1}
T=(\Psi,\hat{T}\Psi)\qquad
T^{\nu}_{\mu}=(\Psi,\hat{T}^{\nu}_{\mu}\Psi)
\end{equation}
where $\hat{T}$ is the energy-momentum tensor operator and $\Psi$
is a state vector.

The Hamiltonian
\begin{equation}\label{3.5.2}
\hat{H}_{t}=\int\limits_{S}\eta\;\hat{T}_{0}^{0}
\end{equation}
where
\begin{equation}\label{3.5.3}
\int\limits_{S}\eta=\int\limits_{S}\eta^{\mathrm{space}}:=
\int\limits_{S}\sqrt{|h|}d^{3}x\qquad |h|=\mathrm{det}(h_{ij})
\end{equation}

The next problem in the Hamiltonian formulation of quantum field
theory is this: Being based on the equivalence principle, to
construct the operator $\hat{T}$ for free fields in such a way
that the Hamiltonian be of the form
\begin{equation}\label{3.5.4}
\hat{H}_{t}=\sum_{m}\omega_{m}(t)\hat{a}^{\dag}_{m}\hat{a}_{m}
\end{equation}

\section{Scalar field}

\subsection{Energy-momentum tensor and Hamiltonian}

For a real scalar field $\varphi$, the energy-momentum tensor is
of the form
\begin{equation}\label{4.1.1}
T_{\mu\nu}=\varphi_{,\mu}\varphi_{,\nu}-
\frac{1}{2}g_{\mu\nu}g^{\sigma\lambda}\varphi_{,\sigma}\varphi_{,\lambda}+
\frac{1}{2}g_{\mu\nu}M^{2}\varphi^{2}
\end{equation}
so that for the metric (3.1.3)
\begin{equation}\label{4.1.2}
T_{0}^{0}=\frac{1}{2}g^{00}\varphi_{,0}\varphi_{,0}+\frac{1}{2}h^{jl}
\varphi_{,j}\varphi_{,l}+\frac{1}{2}M^{2}\varphi^{2}
\end{equation}
The Hamiltonian
\begin{equation}\label{4.1.3}
H=\int\eta\:T_{0}^{0}=\frac{1}{2}\int\eta\;[g^{00}\varphi_{,0}\varphi_{,0}-
\varphi\bigtriangleup_{h}\varphi+M^{2}\varphi^{2}]
\end{equation}
where $\bigtriangleup_{h}$ is the Laplacian on $S$:
\begin{equation}\label{4.1.4}
\bigtriangleup_{h}\varphi=\frac{1}{\sqrt{|h|}}[\sqrt{|h|}g^{jl}\varphi_{,j}]_{,l}
\end{equation}

\subsection{Space modes and field operator expansion}

Introduce space modes $f$ as solutions to the equation
\begin{equation}\label{4.2.1}
\bigtriangleup_{h}f=-k^{2}f\qquad f=f(s,t)\qquad k^{2}=k^{2}(t)
\end{equation}
with the conditions
\begin{equation}\label{4.2.2}
\int\eta\:f^{\ast}_{m}f_{m'}=\delta_{mm'}
\end{equation}
\begin{equation}\label{4.2.3}
(k^{2}_{m}-k^{2}_{m'})\int\eta\:f_{m}f_{m'}=0
\end{equation}

For the field operator (in fact, operator-valued distribution), we
put in the Schr\"odinger picture
\begin{equation}\label{4.2.4}
\hat{\varphi}(s,t)=\sum_{m}\frac{1}{\sqrt{2\omega_{m}(t)}}
[f_{m}(s,t)\hat{a}_{m}+f_{m}^{\ast}(s,t)\hat{a}_{m}^{\dagger}]
\end{equation}

\subsection {Inertial time derivation}

Now we switch from time derivatives $(\cdots)_{,0}$ to inertial
time derivatives $(\cdots)_{:0}$:
\begin{equation}\label{4.3.1}
(\cdots)_{,0}\stackrel{\mathrm{switch}}{\longrightarrow}(\cdots)_{:0}
\end{equation}
The inertial time derivation is an implementation of the
equivalence principle by means of an imitation of the derivation
in inertial reference frames in flat spacetime.

For the scalar field (2.2.4) in the Heisenberg picture, the
inertial time derivation is defined by
\begin{equation}\label{4.3.2}
\hat{\varphi}_{:0}=\frac{1}{\sqrt{2}}\sum_{m}
\left[\left(\frac{1}{\sqrt{\omega_{m}}}f_{m}\hat{a}_{m}\right)_{:0}+
\left(\frac{1}{\sqrt{\omega_{m}}}f_{m}^{\ast}\hat{a}_{m}^{\dagger}\right)_{:0}\right]
\end{equation}
\begin{equation}\label{4.3.3}
\left(\frac{1}{\sqrt{\omega_{m}}}f_{m}\hat{a}_{m}\right)_{:0}=
-\mathrm{i}\sqrt{g_{00}}\omega_{m}\left(\frac{1}{\sqrt{\omega_{m}}}f_{m}\hat{a}_{m}\right)
\end{equation}
\begin{equation}\label{4.3.4}
\left(\frac{1}{\sqrt{\omega_{m}}}f_{m}^{\ast}\hat{a}_{m}^{\dagger}\right)_{:0}=
\mathrm{i}\sqrt{g_{00}}\omega_{m}\left(\frac{1}{\sqrt{\omega_{m}}}f_{m}^{\ast}\hat{a}_{m}^{\dagger}\right)
\end{equation}
so that
\begin{equation}\label{4.3.5}
\hat{\varphi}_{:0}=\mathrm{i}\sqrt{g_{00}}\sum_{m}\frac{1}{\sqrt{2\omega_{m}}}
[-\omega_{m}f_{m}\hat{a}_{m}+\omega_{m}f_{m}^{\ast}\hat{a}_{m}^{\dagger}]
\end{equation}

\subsection{The Hamiltonian}

We obtain from (4.1.3), (4.2.4), (4.2.1), (4.3.5)
\begin{eqnarray}
\hat{H}&=&\frac{1}{2}\int\eta\:[g^{00}\hat{\varphi}_{:0}\hat{\varphi}_{:0}-
\hat{\varphi}\triangle_{h}\hat{\varphi}+M^{2}\varphi^{2}]\nonumber\\
&=&\frac{1}{4}\sum_{mm'}\frac{1}{\sqrt{\omega_{m}\omega_{m'}}}\int\eta\:
[(-\omega_{m}\omega_{m'}+k^{2}_{m`}+M^{2})(f_{m}f_{m'}\hat{a}_{m}\hat{a}_{m'}+
f_{m}^{\ast}f_{m'}^{\ast}\hat{a}_{m}^{\dagger}\hat{a}_{m'}^{\dagger})\nonumber\\
&{}&\qquad\qquad\qquad\qquad+(\omega_{m}\omega_{m'}+k^{2}_{m'}+M^{2})
(f_{m}f_{m'}^{\ast}\hat{a}_{m}\hat{a}_{m'}^{\dagger}+
f_{m}^{\ast}f_{m'}\hat{a}_{m}^{\dagger}\hat{a}_{m'}]
\end{eqnarray}
Put
\begin{equation}\label{4.4.2}
\omega_{m}=\sqrt{M^{2}+k^{2}_{m}}
\end{equation}
then
\begin{equation}\label{4.4.3}
\hat{H}=\frac{1}{2}\sum_{m}\omega_{m}(\hat{a}_{m}\hat{a}_{m}^{\dagger}+
\hat{a}_{m}^{\dagger}\hat{a}_{m})
\end{equation}
Normal ordering produces in the Schr\"odinger picture
\begin{equation}\label{4.4.4}
\hat{H}_{S}(t)=\sum_{m}\omega_{m}\hat{a}_{m}^{\dagger}\hat{a}_{m}\qquad
\hat{a}_{m}=\hat{a}_{mS}
\end{equation}
In the Heisenberg picture,
\begin{equation}\label{4.4.5}
\hat{a}_{mH}(t)=\mathrm{e}^{-\mathrm{i}\beta_{m}(t)}\hat{a}_{m}\qquad
\hat{a}_{mH}^{\dagger}(t)=\mathrm{e}^{\mathrm{i}\beta_{m}(t)}\hat{a}_{m}^{\dagger}
\qquad \beta(t)=\int\limits_{0}^{t}\omega_{m}(t)dt
\end{equation}
and
\begin{equation}\label{4.4.6}
\hat{H}_{H}(t)=\hat{H}_{S}(t)
\end{equation}

\subsection{Charged scalar field}

The energy-momentum tensor operator is
\begin{equation}\label{4.5.1}
\hat{T}_{\mu\nu}=:\hat{\varphi}_{,\mu}^{\dagger}\hat{\varphi}_{,\nu}+
\hat{\varphi}_{,\nu}^{\dagger}\hat{\varphi}_{,\mu}-
g_{\mu\nu}g^{\sigma\lambda}\hat{\varphi}_{,\sigma}^{\dagger}\hat{\varphi}_{,\lambda}+
g_{\mu\nu}M^{2}\hat{\varphi}^{\dagger}\hat{\varphi}:
\end{equation}
The Hamiltonian
\begin{equation}\label{4.5.2}
\hat{H}=\int\eta\:
:[g^{00}\hat{\varphi}_{:0}^{\dagger}\hat{\varphi}_{:0}-
\hat{\varphi}^{\dagger}\bigtriangleup_{h}\hat{\varphi}+M^{2}\hat{\varphi}^{\dagger}\hat{\varphi}]:
\end{equation}
The field operator
\begin{equation}\label{4.5.3}
\hat{\varphi}=\sum_{m}\frac{1}{\sqrt{2\omega_{m}}}[f_{m}\hat{a}_{(+)m}+
f_{m}^{\ast}\hat{a}_{(-)m}^{\ast}]
\end{equation}
\begin{equation}\label{4.5.4}
\hat{\varphi}^{\dagger}=\sum_{m}\frac{1}{\sqrt{2\omega_{m}}}[f_{m}^{\ast}\hat{a}_{(+)m}^{\dagger}+
f_{m}\hat{a}_{(-)m}]
\end{equation}
We obtain
\begin{equation}\label{4.5.5}
\hat{H}_{H}(t)=\hat{H}_{S}(t)=\sum_{m}\omega_{m}(t)[\hat{a}_{(-)m}^{\dagger}\hat{a}_{(-)m}+
\hat{a}_{(+)m}^{\dagger}\hat{a}_{(+)m}]
\end{equation}

\section{Vector field}

\subsection{Energy-momentum tensor and Hamiltonian}

For a massive vector field $A$, the energy-momentum tensor is of
the form
\begin{equation}\label{5.1.1}
T_{\mu\nu}=\frac{1}{4}g_{\mu\nu}F_{\lambda\sigma}F^{\lambda\sigma}-
F_{\mu}{}^{\lambda}F_{\nu\lambda}+M^{2}(A_{\mu}A_{\nu}-
\frac{1}{2}g_{\mu\nu}A_{\lambda}A^{\lambda})
\end{equation}
where
\begin{equation}\label{5.1.2}
F_{\mu\nu}=A_{\nu;\mu}-A_{\mu;\nu}=A_{\nu,\mu}-A_{\mu,\nu}
\end{equation}
so that for the metric (3.1.3)
\begin{equation}\label{5.1.3}
T_{0}^{0}=\frac{1}{2}[A_{j,l}F^{lj}-M^{2}A_{j}A^{j}]-
\frac{1}{2}[A_{0,l}F^{l0}-M^{2}A_{0}A^{0}]-\frac{1}{2}A_{j,0}F^{0j}
\end{equation}
The Hamiltonian
\begin{equation}\label{5.1.4}
H=\int\eta\:T_{0}^{0}
\end{equation}

We have
\begin{equation}\label{5.1.5}
\int\eta\:A_{j,l}F^{lj}=-\int\eta\:A_{j}\frac{1}{\sqrt{|h|}}(\sqrt{|h|}F^{lj})_{,l}
\end{equation}
and
\begin{equation}\label{5.1.6}
\int\eta\:A_{0,l}F^{l0}=-\int\eta\:A_{0}\frac{1}{\sqrt{|h|}}(\sqrt{|h|}F^{l0})_{,l}
\end{equation}
With the standard metric (3.1.3),
$(1/\sqrt{|h|})(\sqrt{|h|})F^{lj})_{,l}$ is a 3-vector and
$(1/\sqrt{|h|})(\sqrt{|h|})F^{l0})_{,l}$ is a scalar, so that
$(1/\sqrt{|h|})(\sqrt{|h|})F^{l\mu})_{,l}$ makes sense.

\subsection{Field operator expansion}

Put
\begin{equation}\label{5.2.1}
\hat{A}=\sum_{m}\frac{1}{\sqrt{2\omega_{m}}}\sum_{n=1}^{3}[f_{m}e_{mn}\hat{a}_{mn}+
f_{m}^{\ast}e_{mn}^{\ast}\hat{a}_{mn}^{\dagger}]
\end{equation}
where the space modes $f_{m}$ are defined as in Subsection 4.2,
and
\begin{equation}\label{5.2.2}
e_{mn}=e_{mn}(s,t)\qquad n=1,2,3
\end{equation}
are the polarization vectors; in components
\begin{equation}\label{5.2.3}
\hat{A}_{\mu}=\sum_{m}\frac{1}{\sqrt{2\omega_{m}}}\sum_{n=1}^{3}[f_{m}e_{mn\mu}\hat{a}_{mn}+
f_{m}^{\ast}e_{mn\mu}^{\ast}\hat{a}_{mn}^{\dagger}]
\end{equation}
with orthonormalization conditions
\begin{equation}\label{5.2.4}
e_{mn\mu}e_{mn'}^{\ast\mu}=-\delta_{nn'}
\end{equation}

\subsection{Inertial time derivation}

Introduce inertial time derivatives:
\begin{equation}\label{5.3.1}
\left(\frac{1}{\sqrt{\omega_{m}}}f_{m}e_{mn}\hat{a}_{mn}\right)_{:0}=
-\mathrm{i}\sqrt{g_{00}}\omega_{m}\left(\frac{1}{\sqrt{\omega_{m}}}f_{m}e_{mn}\hat{a}_{mn}\right)
\end{equation}
\begin{equation}\label{5.3.2}
\left(\frac{1}{\sqrt{\omega_{m}}}f_{m}^{\ast}e_{mn}^{\ast}\hat{a}_{mn}^{\dagger}\right)_{:0}=
\mathrm{i}\sqrt{g_{00}}\omega_{m}\left(\frac{1}{\sqrt{\omega_{m}}}
f_{m}^{\ast}e_{mn}^{\ast}\hat{a}_{mn}^{\dagger}\right)
\end{equation}
\begin{equation}\label{5.3.3}
\left(\frac{1}{\sqrt{\omega_{m}}}f_{m}e_{mn}\hat{a}_{mn}\right)_{:00}=
-g_{00}\omega_{m}^{2}\left(\frac{1}{\sqrt{\omega_{m}}}f_{m}e_{mn}\hat{a}_{mn}\right)
\end{equation}
\begin{equation}\label{5.3.4}
\left(\frac{1}{\sqrt{\omega_{m}}}f_{m}^{\ast}e_{mn}^{\ast}\hat{a}_{mn}^{\dagger}\right)_{:00}=
-g_{00}\omega_{m}^{2}\left(\frac{1}{\sqrt{\omega_{m}}}
f_{m}^{\ast}e_{mn}^{\ast}\hat{a}_{mn}^{\dagger}\right)
\end{equation}

In flat spacetime, the equations
\begin{equation}\label{5.3.5}
F^{\nu\mu}{}{}_{,\nu}+M^{2}A^{\mu}=0
\end{equation}
and
\begin{equation}\label{5.3.6}
A^{\nu}{}_{,\nu}=0
\end{equation}
are fulfilled. So we put
\begin{equation}\label{5.3.7}
F^{0\mu}{}{}_{:0}+\frac{1}{\sqrt{|h|}}(\sqrt{|h|}F^{l\mu})_{,l}+M^{2}A^{\mu}=0
\end{equation}
and
\begin{equation}\label{5.3.8}
A^{0}{}_{:0}+\frac{1}{\sqrt{|h|}}(\sqrt{|h|}A^{l})_{,l}=0
\end{equation}

Finally, we put
\begin{equation}\label{5.3.9}
[(\cdots)_{,l}]_{:0}=:(\cdots)_{,l:0}=(\cdots)_{:0,l}:=[(\cdots)_{:0}]_{,l}
\end{equation}

\subsection{The Hamiltonian and constraints on the polarization vectors}

The Hamiltonian (5.1.4) works out to be
\begin{equation}\label{5.4.1}
H=\frac{1}{2}\int\eta\:\{g^{00}[A^{\mu}(A_{\mu})_{:00}-
(A^{\mu})_{:0}(A_{\mu})_{:0}]+g^{00}{}{}_{,l}[A^{l}(A_{0})_{:0}-(A^{l})_{:0}A_{0}]\}
\end{equation}
Now \begin{eqnarray}\!\!\!\!\!\!\!\!
g^{00}:[\hat{A}^{\mu}(\hat{A}_{\mu})_{:00}-
(\hat{A}^{\mu})_{:0}(\hat{A}_{\mu})_{:0}]:\!\!&=&\!\!
\frac{1}{2}\sum_{mm'}\left(\frac{\omega_{m'}}{\omega_{m}}\right)^{1/2}\nonumber\\
&{}&\!\!\!\!\!\times\sum_{nn'}\{-([(\omega_{m}+\omega_{m'})f^{\ast}_{m}f_{m'}
e_{mn}^{\ast\mu}e_{m'n'\mu}\hat{a}_{mn}^{\dagger}\hat{a}_{m'n'}]
+[\cdots]^{\dagger})\nonumber\\
&{}&+([(\omega_{m}-\omega_{m'})f_{m}f_{m'}
e_{mn}^{mu}e_{m'n'\mu}\hat{a}_{mn}\hat{a}_{m'n'}]
+[\cdots]^{\dagger})\}
\end{eqnarray}
and
\begin{eqnarray}
:[\hat{A}^{l}(\hat{A}_{0})_{:0}-(\hat{A}^{l})_{:0}\hat{A}_{0}]:&=&
\frac{\mathrm{i}}{2}\sqrt{g_{00}}
\sum_{mm'}\frac{1}{\sqrt{\omega_{m}\omega_{m'}}}\nonumber\\
&{}&\times\sum_{nn'}\{[(\omega_{m'}-\omega_{m})f_{m'}f_{m}e_{m'n'}^{l}e_{mn0}
\hat{a}_{m'n'}\hat{a}_{mn}\nonumber\\
&{}&+(\omega_{m'}+\omega_{m})f_{m'}f_{m}^{\ast}e_{m'n'}^{l}e_{mn0}^{\ast}
\hat{a}_{mn}^{\dagger}\hat{a}_{m'n'}]-[\cdots]^{\dagger}\}
\end{eqnarray}

Impose constraints on the $(e_{mn})$:
\begin{equation}\label{5.4.4}
e_{mn\mu}e_{mn'}^{\ast\mu}=-\delta_{nn'}
\end{equation}
\begin{equation}\label{5.4.5}
-\mathrm{i}\sqrt{g_{00}}\omega_{m}f_{m}e_{mn}^{0}+
\frac{1}{\sqrt{|h|}}(\sqrt{|h|}f_{m}e_{mn}^{l})_{,l}=0
\end{equation}
\begin{equation}\label{5.4.6}
(\omega_{m'}-\omega_{m})\int\eta\:\sqrt{g_{00}}g^{00}{}{}_{,l}f_{m'}f_{m}
e_{m'n'}^{l}e_{mn0}=0
\end{equation}
\begin{equation}\label{5.4.7}
\int\eta\:\sqrt{g_{00}}g^{00}{}{}_{,l}f_{m'}f_{m}^{\ast}
e_{m'n'}^{l}e_{mn0}^{\ast}=0
\end{equation}
\begin{equation}\label{5.4.8}
(\omega_{m}-\omega_{m'})\int\eta\:f_{m}f_{m'}
e_{mn}^{\mu}e_{m'n'\mu}=0
\end{equation}
\begin{equation}\label{5.4.9}
\int\eta\:f_{m}^{\ast}f_{m'} e_{mn}^{\ast\mu}e_{m'n'\mu}=0\quad
\mathrm{for} \;m'\neq m
\end{equation}
\begin{equation}\label{5.4.10}
\int\eta\:f_{m}^{\ast}f_{m}=1
\end{equation}
Equations (5.4.9), (5.4.10) and (5.4.4) imply
\begin{equation}\label{5.4.11}
\int\eta\:f_{m}f_{m'}
e_{mn}^{\mu}e_{m'n'\mu}=-\delta_{mm'}\delta_{nn'}
\end{equation}
Then the Hamiltonian
\begin{equation}\label{5.4.12}
\hat{H}_{H}(t)=\hat{H}_{S}(t)=
\sum_{m}\omega_{m}(t)\sum_{n}\hat{a}_{mn}^{\dagger}\hat{a}_{mn}
\end{equation}

\subsection{Massless vector field}

We use the radiation gauge:
\begin{equation}\label{5.5.1}
\hat{A}_{0}=0
\end{equation}
Then $n=1,2$ and
\begin{equation}\label{5.5.2}
e_{mn}^{0}=0
\end{equation}

\section{Dirac field}

\subsection{Flat spacetime}

In flat spacetime, the energy-momentum tensor is of the form
\begin{equation}\label{6.1.1}
T_{\alpha\beta}=\frac{1}{4}
\{[\bar{\psi}(\gamma_{(\alpha}\mathrm{i}\partial_{\beta)})\psi]+
[\cdots]^{\dagger}\}\qquad \alpha,\beta=0,1,2,3
\end{equation}
Here $\gamma^{\alpha}$ are Dirac matrices that satisfy the
anticommutative relations
\begin{equation}\label{6.1.2}
\{\gamma^{\alpha},\gamma^{\beta}\}=2\eta^{\alpha\beta}\qquad
\eta=\mathrm{diag}(1,-1,-1,-1)\qquad\gamma^{0\dagger}=\gamma^{0}
\quad\gamma^{a\dagger}=-\gamma^{a}\quad a=1,2,3
\end{equation}
\begin{equation}\label{6.1.3}
(\alpha\cdots\beta)=\alpha\cdots\beta+\beta\cdots\alpha
\end{equation}
\begin{equation}\label{6.1.4}
\bar{\psi}=\psi^{\dagger}\gamma^{0}
\end{equation}
The Dirac equation reads
\begin{equation}\label{6.1.5}
\mathrm{i}\gamma^{\alpha}\psi_{,\alpha}-M\psi=0
\end{equation}

\subsection{Curved spacetime}

The curved spacetime generalization of the above formulas is given
by the following replacements [2]:
\begin{equation}\label{6.2.1}
\gamma^{\alpha}\rightarrow
\gamma^{\mu}=V_{\alpha}^{\mu}\gamma^{\alpha}\qquad \mu=0,1,2,3
\end{equation}
\begin{equation}\label{6.2.2}
\partial_{\alpha}\rightarrow \triangledown_{\mu}=
\partial_{\mu}+\Gamma_{\mu}
\end{equation}
where
\begin{equation}\label{6.2.3}
\{\gamma^{\mu},\gamma^{\nu}\}=2g^{\mu\nu}\qquad
\end{equation}
\begin{equation}\label{6.2.4}
\Gamma_{\mu}=\frac{1}{2}\Sigma^{\alpha\beta}V_{\alpha}^{\nu}V_{\beta\nu,\mu}
\qquad\Sigma^{\alpha\beta}=\frac{1}{4}[\gamma^{\alpha},\gamma^{\beta}]
\end{equation}
\begin{equation}\label{6.2.5}
V_{\alpha}^{\mu}V_{\beta}^{\nu}g_{\mu\nu}=\eta_{\alpha\beta}
\end{equation}
\begin{equation}\label{6.2.6}
V_{\alpha}^{\mu}=V_{(\alpha)}^{\mu}\qquad
\{V_{(\alpha)}:\alpha=0,1,2,3\} \;\mathrm{is\;a\;tetrad}
\end{equation}

Now
\begin{equation}\label{6.2.7}
T_{\mu\nu}=\frac{1}{4}
\{[\bar{\psi}(\gamma_{(\mu}\mathrm{i}\partial_{\nu)})\psi]+
[\cdots]^{\dagger}\}
\end{equation}
and
\begin{equation}\label{6.2.8}
\mathrm{i}\gamma^{\mu}\triangledown_{\mu}\psi-M\psi=0
\end{equation}

Write (6.2.7) as
\begin{equation}\label{6.2.9}
T_{\mu\nu}=\psi^{\dagger}\frac{1}{4}(K_{\mu\nu}+K_{\mu\nu}^{\dagger})\psi
\end{equation}
where
\begin{equation}\label{6.2.10}
K_{\mu\nu}=\gamma^{0}[\gamma_{(\mu}\mathrm{i}\triangledown_{\nu)}]=K_{\nu\mu}
\end{equation}
specifically,
\begin{equation}\label{6.2.11}
K_{00}=2\gamma^{0}\gamma_{0}\mathrm{i}\triangledown_{0}
\end{equation}

\subsection{Standard metric}

We now turn to the case of the standard metric (3.1.3). Put
\begin{equation}\label{6.3.1}
V_{(0)}^{j}=0\quad j=1,2,3
\end{equation}
Then
\begin{equation}\label{6.3.2}
\gamma^{0}=V_{(0)}^{0}\gamma^{0}_{D}\qquad\gamma^{0}_{D}
:=\gamma^{0}_{\mathrm{Dirac}}=\gamma^{\alpha=0}\qquad
[V^{0}_{(0)}]^{2}=g^{00}
\end{equation}
The Hamiltonian
\begin{equation}\label{6.3.3}
H=\int\eta\:T_{0}^{0}
\end{equation}
so that we are interested mainly in
\begin{equation}\label{6.3.4}
T_{0}^{0}=g^{00}T_{00}=
g^{00}\psi^{\dagger}\frac{1}{4}(K_{00}+K_{00}^{\dagger})\psi
\end{equation}
with $K_{00}$ given by (6.2.11).

\subsection{Inertial time derivation}

Introduce an inertial time derivative:
\begin{equation}\label{6.4.1}
\triangledown_{0}\stackrel{\mathrm{switch}}{\longrightarrow}\triangledown_{:0}
\end{equation}
Since
\begin{equation}\label{6.4.2}
\triangledown_{0}=\partial_{0}+\Gamma_{0}
\end{equation}
we consider
\begin{equation}\label{6.4.3}
\partial_{0}\stackrel{\mathrm{switch}}{\longrightarrow}\partial_{:0}
\end{equation}
In the spirit of the inertial time derivation, we take
\begin{equation}\label{6.4.4}
\partial_{:0}g^{00}=g^{00}\partial_{:0}\qquad
\triangledown_{:0}g^{00}=g^{00}\triangledown_{:0}
\end{equation}
and, accordingly, put
\begin{equation}\label{6.4.5}
T_{0}^{0}=\psi^{\dagger}\Theta_{0}^{0}\psi
\end{equation}
where
\begin{equation}\label{6.4.6}
\Theta_{0}^{0}=\frac{1}{4}[(g^{00}K_{00})+(g^{00}K_{00})^{\dagger}]
\qquad [\Theta_{0}^{0}]^{\dagger}=\Theta_{0}^{0}
\end{equation}
\begin{equation}\label{6.4.7}
K_{00}=2\gamma^{0}\gamma_{0}\mathrm{i}\triangledown_{:0}
\end{equation}

The next step is as follows. With (6.2.8) in mind, we put
\begin{equation}\label{6.4.8}
\mathrm{i}\triangledown_{:0}=
(\gamma^{0})^{-1}[\gamma^{l}(-\mathrm{i}\triangledown_{l})+M]
\end{equation}
so that
\begin{equation}\label{6.4.9}
K_{00}=2\gamma^{0}\gamma_{0}(\gamma^{0})^{-1}
[\gamma^{l}(-\mathrm{i}\triangledown_{l})+M]
\end{equation}
Now,
\begin{equation}\label{6.4.10}
\gamma^{0}\gamma_{0}(\gamma^{0})^{-1}=V_{(0)}^{0}g_{00}\gamma^{0}_{D}
\end{equation}
and
\begin{equation}\label{6.4.11}
g^{00}K_{00}=2V_{(0)}^{0}\gamma^{0}_{D}
[\gamma^{l}(-\mathrm{i}\triangledown_{l})+M]
\end{equation}

Note that with (6.4.8) the relation (6.4.4) does not hold; it was
introduced only to arrive at the expression (6.4.6).

We obtain
\begin{equation}\label{6.4.12}
(g^{00}K_{00})^{\dagger}=2\gamma^{0}_{D}
[\gamma^{l}(-\mathrm{i}\triangledown_{l})+M]V_{(0)}^{0}
\end{equation}
so that finally
\begin{equation}\label{6.4.13}
\Theta_{0}^{0}=\frac{1}{2}\gamma^{0}_{D}[\{V_{(0)}^{0}\gamma^{l},
-\mathrm{i}\triangledown_{l}\}+2V_{(0)}^{0}M]\qquad[V_{(0)}^{0}]^{2}=g^{00}
\end{equation}
where $\{\cdots,\cdots\}$ is an anticommutator.

\subsection{Space modes and field operator}

Introduce space modes by the equation
\begin{equation}\label{6.5.1}
\Theta_{0}^{0}f_{m}=E_{m}f_{m}\qquad f_{m}=f_{m}(s,t)\qquad
E_{m}=E_{m}(t)
\end{equation}
and orthonormalize them according to
\begin{equation}\label{6.5.2}
\int\eta\:f_{m'}^{\dagger}f_{m}=2\omega_{m}\delta_{m'm}
\end{equation}
where
\begin{equation}\label{6.5.3}
\omega_{m}=|E_{m}|\qquad E_{m}\neq 0
\end{equation}

Put
\begin{equation}\label{6.5.4}
\hat{\psi}=\sum_{m}^{E_{m}\neq
0}\frac{1}{\sqrt{2\omega_{m}}}f_{m}[\theta(E_{m})\hat{a}_{m}+
\theta(-E_{m})\hat{b}_{m}^{\dagger}]
\end{equation}
\begin{equation}\label{6.5.5}
\hat{\psi}^{\dagger}=\sum_{m}^{E_{m}\neq
0}\frac{1}{\sqrt{2\omega_{m}}}f_{m}^{\dagger}[\theta(E_{m})\hat{a}_{m}^{\dagger}+
\theta(-E_{m})\hat{b}_{m}]
\end{equation}

\subsection{The energy-momentum tensor operator and the Hamiltonian}

The energy-momentum tensor operator is
\begin{equation}\label{6.6.1}
\hat{T}_{\mu\nu}=:\hat{\psi}^{\dagger}\frac{1}{4}(K_{\mu\nu}+
K_{\mu\nu}^{\dagger})\hat{\psi}:
\end{equation}
where $K_{\mu\nu}$ is given by (6.2.10) with the switch (6.4.1),
(6.4.8).

The Hamiltonian
\begin{equation}\label{6.6.2}
\hat{H}=\int\eta\::\hat{T}_{0}^{0}:=
\int\eta\::\hat{\psi}^{\dagger}\Theta_{0}^{0}\hat{\psi}:
\end{equation}
with $\Theta_{0}^{0}$ given by (6.4.13). We have
\begin{equation}\label{6.6.3}
\Theta_{0}^{0}f_{m}\theta(\pm E_{m})=\theta(\pm E_{m})E_{m}f_{m}=
\theta(\pm E_{m})(\pm \omega_{m})f_{m}
\end{equation}
so that
\begin{equation}\label{6.6.4}
\hat{H}=\sum_{m}\omega_{m}
[\theta(E_{m})\hat{a}_{m}^{\dagger}\hat{a}_{m}+
\theta(-E_{m})\hat{b}_{m}^{\dagger}\hat{b}_{m}]
\end{equation}
\begin{equation}\label{6.6.5}
E_{m}=E_{m}(t)\qquad \omega_{m}=\omega_{m}(t)\qquad
\hat{H}=\hat{H}_{t}
\end{equation}
\begin{equation}\label{6.6.6}
\hat{H}_{H}=\hat{H}_{S}=\hat{H}_{t}
\end{equation}

\subsection{Reductive reference frame}

In a reductive reference frame, we have
\begin{equation}\label{6.7.1}
g_{00}=1\qquad g^{00}=1\qquad V_{(0)}^{0}=1
\end{equation}
so that
\begin{equation}\label{6.7.2}
\Theta_{0}^{0}=\gamma^{0}_{D}
[\frac{1}{2}\{\gamma^{l},-\mathrm{i}\triangledown_{l}\}+M]
\end{equation}
and
\begin{equation}\label{6.7.3}
[\Theta_{0}^{0}]^{\dagger}=
[-\frac{1}{2}\{\gamma^{l},-\mathrm{i}\triangledown_{l}\}+
M]\gamma^{0}_{D}=\Theta_{0}^{0}
\end{equation}

Consider
\begin{equation}\label{6.7.4}
\Theta_{0}^{0}\Theta_{0}^{0}=[\Theta_{0}^{0}]^{\dagger}\Theta_{0}^{0}=
-\frac{1}{4}\{\gamma^{j},-\mathrm{i}\triangledown_{j}\}
\{\gamma^{l},-\mathrm{i}\triangledown_{l}\}+M^{2}
\end{equation}
Introduce
\begin{equation}\label{6.7.5}
\mathcal{P}:=\frac{1}{2}\{\gamma^{l},-\mathrm{i}\triangledown_{l}\}
\qquad \mathcal{P}^{\dagger}=-\mathcal{P}
\end{equation}
Thus
\begin{equation}\label{6.7.6}
[\Theta_{0}^{0}]^{2}=\mathcal{P}^{\dagger}\mathcal{P}+M^{2}
\end{equation}
We have
\begin{equation}\label{6.7.7}
[\Theta_{0}^{0}]^{2}f_{m}=E_{m}^{2}f_{m}
\end{equation}
or
\begin{equation}\label{6.7.8}
\mathcal{P}^{\dagger}\mathcal{P}f_{m}=(E_{m}^{2}-M^{2})f_{m}
\end{equation}
which implies
\begin{equation}\label{6.7.9}
E_{m}^{2}-M^{2}\geq 0
\end{equation}
Thus
\begin{equation}\label{6.7.10}
\mathcal{P}^{\dagger}\mathcal{P}f_{m}=k_{m}^{2}f_{m}
\end{equation}
and
\begin{equation}\label{6.7.11}
E_{m}^{2}=M^{2}+k_{m}^{2}\qquad E_{m}=\pm \omega_{m}\qquad
\omega_{m}=\sqrt{M^{2}+k_{m}^{2}}
\end{equation}

\subsection{Diagonal space metric}

Let a space metric be diagonal:
\begin{equation}\label{6.8.1}
g^{\mathrm{space}}=\sum_{l}g_{ll}dx^{l}\otimes dx^{l}=
-\sum_{l}h_{ll}dx^{l}\otimes dx^{l}=-h
\end{equation}
Then the vectors $\partial/\partial x^{j},\,\,j=1,2,3,$ are
mutually orthogonal:
\begin{equation}\label{6.8.2}
h\left(\frac{\partial}{\partial x^{j}},\frac{\partial}{\partial
x^{j'} }\right)=\delta_{jj'}h_{jj}
\end{equation}
and it is expedient to choose tetrad vectors along them:
\begin{equation}\label{6.8.3}
(V_{(0)}:a=1,2,3)=(V_{(j)}:j=1,2,3)
\end{equation}
\begin{equation}\label{6.8.4}
V_{(j)}=V_{(j)}^{j}\frac{\partial}{\partial
x^{j}}\;\;(\mathrm{no}\;\sum_{j})\qquad [V_{(j)}^{j}]^{2}=h^{jj}
\qquad h(V_{(j)},V_{(j')})=\delta_{jj'}
\end{equation}
Now we have
\begin{equation}\label{6.8.5}
(V_{(\alpha)}:\alpha=0,1,2,3)=(V_{(\mu)}:\mu=0,1,2,3)
\end{equation}
\begin{equation}\label{6.8.6}
V_{(\mu)}=V_{(\mu)}^{\mu}\frac{\partial}{\partial
x^{\mu}}\;\;(\mathrm{no}\;\sum_{\mu})\qquad
[V_{(\mu)}^{\mu}]^{2}=|g^{\mu\mu}|^{2}\qquad
g(V_{(\mu)},V_{(\mu')})=\eta_{\mu\mu'}
\end{equation}
From (6.2.4) follows
\begin{equation}\label{6.8.7}
\Gamma_{\mu}=0
\end{equation}
so that
\begin{equation}\label{6.8.8}
\triangledown_{\mu}=\partial_{\mu}\qquad\partial_{0}\stackrel{\mathrm{switch}}{\longrightarrow}\partial_{:0}
\end{equation}

Next,
\begin{equation}\label{6.8.9}
\gamma^{l}=V_{(l)}^{l}\gamma^{l}_{D}\;\;(\mathrm{no}\;\sum_{l})\qquad
[V_{(l)}^{l}]^{2}=h^{ll}
\end{equation}
For a reductive reference frame with a diagonal metric,
\begin{equation}\label{6.8.10}
\Theta_{0}^{0}=\gamma_{D}^{0}(\mathcal{P}+M)
\end{equation}
\begin{equation}\label{6.8.11}
\mathcal{P}=\pm
\frac{1}{2}\sum_{l}\gamma^{l}_{D}\{\sqrt{h^{ll}},-\mathrm{i}\partial_{l}\}
\end{equation}

\section{Massless Weyl field}

\subsection{Flat spacetime}

Let us briefly review the massless Weyl field. In flat spacetime,
the energy-momentum tensor is of the form
\begin{equation}\label{7.1.1}
T_{\alpha\beta}=\sum_{H}^{R,L}T_{H\alpha\beta}
\end{equation}
\begin{equation}\label{7.1.2}
T_{H\alpha\beta}=\frac{1}{2}\psi_{H}^{\dagger}
\{[\sigma_{(\alpha}P_{H}\mathrm{i}\partial_{\beta)}]+[\cdots]^{\dagger}\}\psi_{H}
\end{equation}
Here
\begin{equation}\label{7.1.3}
\mathrm{Hand}=:H=R,L:=\mathrm{Right,\;Left}
\end{equation}
$\sigma^{a},\;a=1,2,3,$ are the Pauli matrices, $\sigma^{0}=I$,
\begin{equation}\label{7.1.4}
P_{L}=1\qquad
P_{R}\mathrm{i}\partial_{0}=\mathrm{i}\partial_{0}\qquad
P_{R}\mathrm{i}\partial_{a}=-\mathrm{i}\partial_{a}
\end{equation}
The equation
\begin{equation}\label{7.1.5}
\sigma^{\alpha}(P_{H}\mathrm{i}\partial_{\alpha})\psi_{H}=0
\end{equation}
holds.

\subsection{Curved spacetime}

The energy-momentum tensor
\begin{equation}\label{7.2.1}
T_{\mu\nu}=\sum_{H}^{R,L}T_{H\mu\nu}
\end{equation}
\begin{equation}\label{7.2.2}
T_{H\mu\nu}=\psi_{H}^{\dagger}\frac{1}{2}(K_{H\mu\nu}+
K_{H\mu\nu}^{\dagger})\psi_{H}
\end{equation}
where
\begin{equation}\label{7.2.3}
K_{H\mu\nu}=\sigma_{(\mu}P_{H}\mathrm{i}\triangledown_{\nu)}
\end{equation}
Equation (7.1.5) is replaced with
\begin{equation}\label{7.2.4}
\sigma^{\mu}(P_{H}\mathrm{i}\triangledown_{\mu})\psi_{H}=0
\end{equation}

\subsection{Standard metric and inertial time derivation}

The switch from $\triangledown_{0}$ to $\triangledown_{:0}$ is
\begin{equation}\label{7.3.1}
P_{H}\mathrm{i}\triangledown_{0}\stackrel{\mathrm{switch}}{\longrightarrow}
P_{H}\mathrm{i}\triangledown_{:0}=
(\sigma^{0})^{-1}\sigma^{l}P_{H}(-\mathrm{i}\triangledown_{l})
\end{equation}
The Hamiltonian
\begin{equation}\label{7.3.2}
H=\sum_{H}^{R,L}H_{H}
\end{equation}
\begin{equation}\label{7.3.3}
H_{H}=\int\eta\:T_{H0}^{0}
\end{equation}
and
\begin{equation}\label{7.3.4}
T_{H0}^{0}=\psi_{H}^{\dagger}\Theta_{H0}^{0}\psi_{H}
\end{equation}
where
\begin{equation}\label{7.3.5}
\Theta_{H0}^{0}=\frac{1}{2}[(g^{00}K_{H00})+(g^{00}K_{H00})^{\dagger}]
\end{equation}
\begin{equation}\label{7.3.6}
K_{H00}=\sigma_{(0}P_{H}\mathrm{i}\triangledown_{:0)}=
2\sigma_{0}(\sigma^{0})^{-1}\sigma^{l}P_{H}(-\mathrm{i}\triangledown_{l})
\end{equation}

\subsection{Space modes, field operators, and Hamiltonian}

The equation for space modes is
\begin{equation}\label{7.4.1}
\Theta_{H0}^{0}f_{Hm}=E_{Hm}f_{Hm}\qquad \omega_{Hm}=|E_{Hm}|
\end{equation}

Field operators are
\begin{equation}\label{7.4.2}
\hat{\psi}_{H}=\sum_{m}^{E_{Hm}\neq
0}\frac{1}{\sqrt{\omega_{Hm}}}f_{Hm}[\theta(E_{Hm})\hat{a}_{Hm}+
\theta(-E_{Hm})\hat{b}_{\bar{H}m}^{\dagger}]
\end{equation}
where
\begin{equation}\label{7.4.3}
\bar{H}=\left\{
\begin{array}{rcl}
L\quad H=R\\
R\quad H=L\\
\end{array}
\right.
\end{equation}

The Hamiltonian is
\begin{equation}\label{7.4.4}
\hat{H}=\sum_{H}^{R,L}\hat{H}_{H}
\end{equation}
\begin{equation}\label{7.4.5}
\hat{H}_{H}=\sum_{m}\omega_{Hm}(\hat{a}_{Hm}^{\dagger}\hat{a}_{Hm}+
\hat{b}_{Hm}^{\dagger}\hat{b}_{Hm})
\end{equation}

\section{On quantum field state vector}

\subsection{The Einstein equation in semiclassical gravity}

In semiclassical gravity, the Einstein equation reads
\begin{equation}\label{8.1.1}
G-\Lambda g=8\mathrm{\pi}\varkappa(\Psi,\hat{T}\Psi)
\end{equation}
where $G$ is the Einstein tensor, $\Lambda$ is the cosmological
constant, $\varkappa$ is the gravitational constant, and $\Psi$ is
a state vector.

\subsection{Constraints on state vector}

We consider a family of quantum fields,
\begin{equation}\label{8.2.1}
\Phi=\{\hat{\varphi},\hat{A},\hat{\psi},\cdots\}
\end{equation}
for a given metric $g$. So (8.1.1) may be written as
\begin{equation}\label{6.2.2}
(G-\Lambda
g)[g]=8\mathrm{\pi}\varkappa(\Psi,\hat{T}[\Phi[g],g]\Psi)
\end{equation}
where $[g]$ means a dependence on metric and its derivatives. This
equation imposes constraints on $\Psi$. Here are two examples.

If $g$ describes a vacuum spacetime, then
$\Psi=\Psi_{\mathrm{vac}}$.

If a spacetime is static, then $\Psi$ is stationary:
$\hat{H}\Psi=E\Psi$.

\section{The universe}

\subsection{The closed universe}

We consider the closed universe. Cosmic space is a three-sphere:
\begin{equation}\label{9.1.1}
S^{\mathrm{space}}=S^{\mathrm{cosmic}}=S^{3}=
\{x_{k}:k=1,2,3,4,\;\sum_{k}x_{k}^{2}=1\}
\end{equation}
Introduce the radius of the universe, $R(t)$. We have for the
space volume
\begin{equation}\label{9.1.2}
V(t)= \int\eta=\int\limits_{S^{3}}|h_{t}|d^{3}x
\end{equation}
Put
\begin{equation}\label{9.1.3}
R:=(V/2\mathrm{\pi}^{2})^{1/3}\qquad R=R(t)
\end{equation}
and
\begin{equation}\label{9.1.4}
h_{t}=R^{2}(t)\varpi_{t}\qquad
\int\eta_{\varpi}=\int\limits_{S^{3}}\sqrt{|\varpi|}d^{3}x=2\mathrm{\pi}^{2}
\end{equation}
Now
\begin{equation}\label{9.1.5}
g=dt\otimes dt-R^{2}\varpi_{t}\qquad
ds^{2}=dt^{2}-R^{2}\varpi_{ij}dx^{i}dx^{j}
\end{equation}
Thus, we have a reductive reference frame.

\subsection{Scalar and vector fields}

The equation for space modes (4.2.1) takes the form
\begin{equation}\label{9.2.1}
\frac{1}{R^{2}}\triangle_{\varpi}f=-k^{2}f
\end{equation}
or
\begin{equation}\label{9.2.2}
\triangle_{\varpi}f=-\tilde{k}^{2}f
\end{equation}
\begin{equation}\label{9.2.3}
k^{2}(t)=\frac{\tilde{k}^{2}_{t}}{R^{2}(t)}
\end{equation}
Thus
\begin{equation}\label{9.2.4}
\omega_{m}(t)=\sqrt{M^{2}+\frac{\tilde{k}^{2}_{mt}}{R^{2}(t)}}
\end{equation}

\subsection{Dirac field}

We have
\begin{equation}\label{9.3.1}
\Theta_{0}^{0}=\gamma^{0}_{D}(\mathcal{P}+M)\qquad \mathcal{P}=
\frac{1}{2}\{\gamma^{l},-\mathrm{i}\triangledown_{l}\}
\end{equation}
where
\begin{equation}\label{9.3.2}
\gamma^{l}=V^{l}_{(a)}\gamma^{a}\qquad
\triangledown_{l}=\partial_{l}+\Gamma_{l}\qquad \Gamma_{l}=
\frac{1}{8}[\gamma^{a},\gamma^{b}]V_{(a)}^{j}V_{(b)j,l}\qquad
a,b=1,2,3
\end{equation}
Now,
\begin{equation}\label{9.3.3}
h^{jn}V_{(a)j}V_{(b)n}=h_{jn}V_{(a)}^{j}V^{n}_{(b)}=\delta_{ab}
\end{equation}
and
\begin{equation}\label{9.3.4}
h_{jn}=R^{2}\varpi_{jn}\qquad h^{jn}=\frac{1}{R^{2}}\varpi^{jn}
\end{equation}
so that
\begin{equation}\label{9.3.5}
V_{(a)}^{l}=\frac{v_{(a)}^{l}}{R}\qquad V_{(a)l}=Rv_{(a)l}\qquad
v_{(a)}^{l}=\varpi^{lj}v_{(a)j}
\end{equation}
Thus,
\begin{equation}\label{9.3.6}
\mathcal{P}=\frac{1}{R}Q
\end{equation}
\begin{equation}\label{9.3.7}
Q=\frac{1}{2}\{v_{(a)}^{l}\gamma^{a},-\mathrm{i}
(\partial_{l}+\frac{1}{8}[\gamma^{a}\gamma^{b}]v^{j}_{(a)}v_{(b)j,l})\}
\end{equation}
Now (6.7.10) reads
\begin{equation}\label{9.3.8}
\frac{1}{R^{2}}Q^{\dagger}Qf_{m}=k_{m}^{2}f_{m}
\end{equation}
or
\begin{equation}\label{9,3,9}
Q^{\dagger}Qf_{m}=\tilde{k}_{m}^{2}f_{m}
\end{equation}
\begin{equation}\label{9.3.10}
k_{m}^{2}(t)=\frac{\tilde{k}_{mt}^{2}}{R^{2}(t)}
\end{equation}
so that
\begin{equation}\label{9.3.11}
E_{m}^{2}(t)=M^{2}+\frac{\tilde{k}_{mt}^{2}}{R^{2}(t)}\qquad
E_{m}=\pm \omega_{m}\qquad
\omega_{m}(t)=\sqrt{M^{2}+\frac{\tilde{k}_{mt}^{2}}{R^{2}(t)}}
\end{equation}

\subsection{Massless Weyl field}

We obtain
\begin{equation}\label{9.4.1}
K_{H00}=\frac{2}{R}v_{(a)}^{l}\sigma^{a}P_{H}(-\mathrm{i}\triangledown_{l})\
\qquad \sigma^{a}=\sigma^{a}_{\mathrm{Pauli}}
\end{equation}
\begin{equation}\label{9.4.2}
\Theta_{H0}^{0}=\frac{1}{R}\Xi_{H}
\end{equation}
where
\begin{equation}\label{9.4.3}
\Xi_{H}=\sigma^{a}\{v_{(a)}^{l},P_{H}(-\mathrm{i}\triangledown_{l})\}
\end{equation}
Now (7.4.1) reads
\begin{equation}\label{9.4.4}
\frac{1}{R}\Xi_{H}f_{Hm}=E_{Hm}f_{m}
\end{equation}
so that
\begin{equation}\label{9.4.5}
E_{Hm}=\pm \omega_{Hm}\qquad
\omega_{Hm}(t)=\frac{\tilde{k}_{Hmt}}{R(t)}\qquad
\tilde{k}_{Hm}\geq 0
\end{equation}

\subsection{The FLRW universe}

Now consider the
Friedmann-Lama$\hat{\mathrm{\i}}$tre-Robertson-Walker model of the
universe. The Robertson-Walker metric is of the form
\begin{equation}\label{9.5.1}
ds^{2}=dt^{2}-R^{2}(t)\left[\frac{dr^{2}}{1-r^{2}}+
r^{2}(d\theta^{2}+\sin^{2}\theta d\varphi^{2})\right]
\end{equation}
It is diagonal with
\begin{equation}\label{9.5.2}
\varpi=\frac{dr^{2}}{1-r^{2}}+ r^{2}(d\theta^{2}+\sin^{2}\theta
d\varphi^{2})
\end{equation}
independent of time.

Here the quantity $\hat{k}_{m}$ is time independent for all
fields.

\subsection{Cosmological redshift}

For all fields, the result
\begin{equation}\label{9.6.1}
\omega_{m}(t)=\sqrt{M^{2}+\frac{\tilde{k}_{m}^{2}}{R^{2}(t)}}
\end{equation}
has been obtained. Specifically, for photons
\begin{equation}\label{9.6.2}
\omega_{m}(t)=\frac{\tilde{k}_{m}}{R(t)}
\end{equation}
which represents the cosmological redshift. What is essential, is
that the result has been obtained in the context of quantum field
theory.

\section{Black hole spacetime}

\subsection{The Schwarzschild metric}

In this section, the Schwarzschild black hole is considered. The
Schwarzschild metric is of the form
\begin{equation}\label{10.1.1}
ds^{2}=(1-r_{S}/r)dt^{2}-\frac{1}{1-r_{S}/r}dr^{2}-r^{2}(d\theta^{2}+\sin^{2}\theta
d\varphi^{2})
\end{equation}
where $r_{s}=2M_{S}$ is the Schwarzschild radius. Introduce
dimensionless quantities:
\begin{equation}\label{10.1.2}
\tilde{r}=\frac{r}{r_{S}}\qquad \tilde{t}=\frac{t}{r_{S}}\qquad
d\tilde{s}^{2}=\frac{ds^{2}}{r^{2}_{S}}
\end{equation}
\begin{equation}\label{10.1.3}
d\tilde{s}^{2}=(1-1/\tilde{r})d\tilde{t}^{2}-\frac{1}{1-1/\tilde{r}}d\tilde{r}^{2}-
\tilde{r}^{2}(d\theta^{2}+\sin^{2}\theta d\varphi^{2})
\end{equation}
This metric is static and diagonal. It has a physical singularity
at $\tilde{r}=0$, which is unavoidable. In addition, it involves a
coordinate singularity at $\tilde{r}=1$, though
$0<\tilde{r}<\infty$. To eliminate the latter, we have to
introduce a synchronous reference frame.

\subsection{Complete synchronous reference frame}

We will use a complete synchronous reference frame [17] which is
defined as follows. Metric is
\begin{equation}\label{10.2.1}
ds^{2}=d\tau^{2}-\mathrm{e}^{\lambda(\tau,\varrho)}d\varrho^{2}-
r^{2}(\tau,\varrho)(d\theta^{2}+\sin^{2}\theta d\varphi^{2})
\end{equation}
where
\begin{equation}\label{10.2.2}
r=\frac{1}{2}r_{S}\left[\left(\frac{\varrho}{r_{S}}\right)^{2}+1\right]
(1+\cos\eta)
\end{equation}
\begin{equation}\label{10.2.3}
\tau=\frac{1}{2}r_{S}\left[\left(\frac{\varrho}{r_{S}}\right)^{2}+
1\right]^{3/2}(\eta+\sin\eta)
\end{equation}
\begin{equation}\label{10.2.4}
\mathrm{e}^{\lambda}=
\frac{1}{4}\left[\left(\frac{\varrho}{r_{S}}\right)^{2}+1\right]
\frac{[2(1+\cos\eta)^{2}+3(\sin\eta)(\eta+\sin\eta)]^{2}}{(1+\cos\eta)^{2}}
\end{equation}
\begin{equation}\label{10.2.5}
-\mathrm{\pi}<\eta<\mathrm{\pi}\qquad -\infty<\varrho<\infty
\end{equation}
and it is implicit that (10.2.3) determines
\begin{equation}\label{10.2.6}
\eta=\eta(\tau,\varrho)
\end{equation}

Since a change
\begin{equation}\label{10.2.7}
(t,r) \stackrel{\mathrm{change}}{\longrightarrow} (\tau,\varrho)
\end{equation}
is made, the variable $\eta$ may be regarded as representing the
variable $t$.

With $-\infty<\varrho<\infty$, the reference frame under
consideration involves two black and two white holes. To reduce
this to only one black and one white hole, we make a change
\begin{equation}\label{10.2.8}
\varrho \stackrel{\mathrm{change}}{\longrightarrow}
\xi=\varrho^{2}
\end{equation}
Introducing dimensionless quantities
\begin{equation}\label{10.2.9}
\tilde{\tau}=\frac{\tau}{r_{S}}\qquad
\tilde{\xi}=\tilde{\varrho}^{2}=\left(\frac{\varrho}{r_{S}}\right)^{2}
\end{equation}
we obtain
\begin{equation}\label{10.2.10}
d\tilde{s}^{2}= d\tilde{\tau}^{2}-\frac{\mathrm{e}^{\lambda}}
{4\tilde{\xi}}d\tilde{\xi}^{2}-
\tilde{r}^{2}(d\theta^{2}+\sin^{2}\theta d\varphi^{2})\qquad
\tilde{r}>0\qquad \tilde{\xi}>0
\end{equation}
\begin{equation}\label{10.2.11}
\tilde{r}=\frac{1}{2}(\tilde{\xi}+1)(1+\cos\eta)
\end{equation}
\begin{equation}\label{10.2.12}
\tilde{\tau}=\frac{1}{2}(\tilde{\xi}+1)^{3/2}(\eta+\sin\eta)
\end{equation}
\begin{equation}\label{10.2.13}
\mathrm{e}^{\lambda}= \frac{1}{4}(\tilde{\xi}+1)
\frac{[2(1+\cos\eta)^{2}+3(\sin\eta)(\eta+\sin\eta)]^{2}}{(1+\cos\eta)^{2}}
\end{equation}
with
\begin{equation}\label{10.2.14}
-\mathrm{\pi}<\eta<\mathrm{\pi}
\end{equation}

In addition to the physical singularity at $\tilde{r}=0$, the
metric (10.2.10) has a coordinate singularity at $\tilde{\xi}=0$,
but the latter is unessential as $\tilde{\xi}=0$ is a boundary
point.

From (10.2.12) and (10.2.14) it follows that
\begin{equation}\label{10.2.15}
\tilde{\xi}>\theta\left(|\tilde{\tau}|-\frac{\mathrm{\pi}}{2}\right)
\left[\left(\frac{2|\tilde{\tau}|}{\mathrm{\pi}}\right)^{2/3}-1\right]
\end{equation}

\subsection{Geometry of spacetime}

Let us analyze the geometry of the spacetime with the metric
(10.1.3), (10.2.10),
\begin{equation}\label{10.3.1}
(1-1/\tilde{r})d\tilde{t}^{2}-
\frac{1}{1-1/\tilde{r}}d\tilde{r}^{2} =
d\tilde{\tau}^{2}-\frac{\mathrm{e}^{\lambda}}
{4\tilde{\xi}}d\tilde{\xi}^{2}
\end{equation}

(1) $ \tilde{r}=\mathrm{const}$

(10.3.1) reduces to
\begin{equation}\label{10.3.2}
d\tilde{t}=(1+1/\tilde{\xi})^{1/2}d\tilde{\tau}\qquad
\tilde{r}\neq 1
\end{equation}
with $\tilde{\xi}=\tilde{\xi}(\tilde{\tau},\tilde{r})$ via
(10.2.11), (10.2.12).

Let
\begin{equation}\label{10.3.3}
\eta=\mathrm{\pi}-\delta\qquad \delta\ll 1\qquad \tilde{r}>1
\end{equation}
then
\begin{equation}\label{10.3.4}
\tilde{r}=\frac{1}{4}(\tilde{\xi}+1)\delta^{2}\qquad \tilde{\tau}=
\frac{1}{2}(\tilde{\xi}+1)^{3/2}(\mathrm{\pi}-\delta^{3}/6)\qquad\xi\gg
1
\end{equation}
\begin{equation}\label{10.3.5}
\tilde{\tau}=\frac{\mathrm{\pi}}{2}(\tilde{\xi}+1)^{3/2}-
\frac{2}{3}\tilde{r}^{3/2}
\end{equation}
and
\begin{equation}\label{10.3.6}
\tilde{t}_{2}-\tilde{t}_{1}=\tilde{\tau}_{2}-\tilde{\tau}_{1}+
\frac{3}{2}\left(\frac{\mathrm{\pi}}{2}\right)^{2/3}
(\tilde{\tau}_{2}^{1/3}-\tilde{\tau}_{1}^{1/3})\qquad\tilde{\tau}\gg
1
\end{equation}

Let
\begin{equation}\label{10.3.7}
\eta\ll 1
\end{equation}
then
\begin{equation}\label{10.3.8}
\tilde{r}=(\tilde{\xi}+1)(1-\eta^{2}/4)\approx\tilde{\xi}+1
\qquad\tilde{\tau}=(\tilde{\xi}+1)^{3/2}\eta
\end{equation}
and for $\tilde{r}>1$
\begin{equation}\label{10.3.9}
\tilde{\xi}=\tilde{r}-1>0\qquad\tilde{t}_{2}-\tilde{t}_{1}=
(1+1/\tilde{\xi})^{1/2}(\tilde{\tau}_{2}-\tilde{\tau}_{1})\qquad
\frac{\tilde{\tau}_{2}-\tilde{\tau}_{1}}{\tilde{r}^{3/2}}\ll 1
\end{equation}

(2) $\tilde{\xi}=\mathrm{const}$

(10.3.1) reduces to
\begin{equation}\label{10.3.10}
d\tilde{t}=\left(\frac{\tilde{r}}{\tilde{r}-1}\right)^{1/2}
\left[1+\frac{\tilde{r}}{\tilde{r}-1}\frac{1}{\tilde{\xi}+1}
\frac{\sin^{2}\eta}{(1+\cos\eta)^{2}}\right]^{1/2}d\tilde{\tau}
\end{equation}

Let
\begin{equation}\label{10.3.11}
\eta\ll 1
\end{equation}
then
\begin{equation}\label{10.3.12}
d\tilde{t}=\left(\frac{\tilde{\xi}+1}{\tilde{\xi}}\right)^{1/2}
\left[1+\frac{1}{4\tilde{\xi}(\tilde{\xi}+1)^{3}}
\tilde{\tau}^{2}\right]d\tilde{\tau}
\end{equation}

Let
\begin{equation}\label{10.3.13}
\tilde{r}=1+w\qquad 0<w\ll 1
\end{equation}
then
\begin{equation}\label{10.3.14}
\tilde{t}=\tilde{t}_{1}+\log\frac{w_{1}}{w}\qquad
w=w_{1}-\frac{1}{2}(\tilde{\xi}+1)^{1/2}\left[1-
\left(\frac{2}{\tilde{\xi}+1}-1\right)^{2}\right]
(\tilde{\tau}-\tilde{\tau}_{1})=:w_{1}-
\frac{w_{1}}{\tilde{\tau}_{0}(\tilde{\xi})}(\tilde{\tau}-\tilde{\tau}_{1})
\end{equation}
so that
\begin{equation}\label{10.3.15}
w=w_{1}\left[1-\frac{\tilde{\tau}-\tilde{\tau}_{1}}
{\tilde{\tau}_{0}(\tilde{\xi})}\right]\qquad
\tilde{t}=\tilde{t}_{1}+\log\frac{1}{1-(\tilde{\tau}-
\tilde{\tau}_{1})/\tilde{\tau}_{0}(\tilde{\xi})}
\end{equation}
\begin{equation}\label{10.3.16}
\tilde{t}\rightarrow\infty\quad \mathrm{for}\;
\tilde{\tau}\rightarrow\tilde{\tau}_{1}+\tilde{\tau}_{0}(\tilde{\xi})
\end{equation}

(3) $\tilde{\tau}=\mathrm{const}$

(10.3.1) reduces to
\begin{equation}\label{10.3.17}
d\tilde{t}=\frac{1}{4}\left(\frac{\tilde{r}}{\tilde{r}-1}\right)^{1/2}
\left[\frac{\tilde{r}}{\tilde{r}-1}-
\frac{\tilde{\xi}+1}{\tilde{\xi}}\right]^{1/2}
\frac{2(1+\cos\eta)^{2}+3(\sin\eta)(\eta+\sin\eta)}{1+\cos\eta}d\tilde{\xi}
\end{equation}

Let
\begin{equation}\label{10.3.18}
\eta\ll 1 \qquad \tilde{r}>1
\end{equation}
then
\begin{equation}\label{10.3.19}
d\tilde{t}=\frac{\tilde{\tau}}{\tilde{\xi}}\left[\frac{\tilde{\xi}+1}
{4\tilde{\xi}(\tilde{\xi}+1)^{2}-\tilde{\tau}^{2}}\right]^{1/2}d\tilde{\xi}
\end{equation}

\subsection{Quantum fields}

For scalar and vector fields, we obtain the Laplacian
\begin{equation}\label{10.4.1}
\tilde{\triangle}_{\tilde{\xi}\theta\varphi}=
\tilde{\triangle}_{\tilde{\xi}}+
\frac{1}{\tilde{r}^{2}}\triangle_{\theta\varphi} \qquad
\tilde{\tau}=\mathrm{const}
\end{equation}
where
\begin{equation}\label{10.4.2}
\tilde{\triangle}_{\tilde{\xi}}f=
\frac{4\sqrt{\tilde{\xi}}}{\tilde{r}^{2}\mathrm{e}^{\lambda/2}}
\frac{\partial}{\partial\tilde{\xi}}
\left[\sqrt{\tilde{\xi}}\frac{\tilde{r}^{2}}{\mathrm{e}^{\lambda/2}}
\frac{\partial f}{\partial\tilde{\xi}}\right]
\end{equation}

Formulas for spin-$1/2$ fields may be obtained straightforwardly.

A state vector complying to vacuum spacetime is the vacuum one:
\begin{equation}\label{10.4.3}
\Psi=\Psi_{\mathrm{vac}}
\end{equation}
There is no ambiguity here since quantum fields are constructed in
a preferred reference frame.

\section{On interaction}

\subsection{Divergences}

There are two primary sources of difficulties in quantum field
theory, specifically in scattering theory, which manifest
themselves in divergences:

(1) Fields as operator-valued distributions rather than functions.

(2) Perturbation theory, specifically the Dyson expansion, the
expansion of the Green functions, and the expansion in
path-integral methods.

A field at a fixed $x$ is not an (unbounded) operator. The
expression $\mathrm{e}^{-\mathrm{i}\hat{H}t}$ makes sense only if
$\hat{H}$ is an (unbounded) selfadjoint  operator. For free
fields, this does not give rise to difficulties, but for
interacting fields the difficulties are well known.

If $\hat{H}$ is an unbounded operator, the expansion of
$\mathrm{e}^{-\mathrm{i}\hat{H}t}$ in the powers of
$(-\mathrm{i}\hat{H}t)$ may be incorrect [18].

\subsection{Cutoff}

The simplest way to transform a field into an operator is to
introduce cutoff for space modes:
\begin{equation}\label{11.2.1}
f_{m}\mapsto f_{m}\zeta(\omega_{m})
\end{equation}
There is a possibility to introduce $\zeta(\omega_{m})$ in a
natural way. Introduce the smooth function
\begin{equation}\label{11.2.2}
\eta(x)=\left\{\begin{array}{lcl} 1\;\;\mathrm{for}\;\;x\leq 1\\
1-\exp\{\mathrm{e}^{-x}/(1-x)\}\;\mathrm{for}\;x>1\\
0\;\;\mathrm{for}\;\;x=\infty
\end{array}
\right.
\end{equation}
and
\begin{equation}\label{11.2.3}
\zeta(\omega)=\eta\left(\frac{\omega^{2}}{1/\varkappa}+
\frac{\varkappa\Lambda^{2}}{\omega^{2}}\right)
\end{equation}
where $\varkappa$ is the gravitational constant
$(\varkappa=t_{\mathrm{P}}^{2},\;t_{\mathrm{P}}\;\mathrm{is\;
the\;Planck\;time})$ and $\Lambda$ is the cosmological constant.
We have
\begin{equation}\label{11.2.4}
\varkappa\Lambda^{2}\lesssim\omega^{2}\lesssim 1/\varkappa
\end{equation}
i.e., both an ultraviolet and an infrared cutoff. For special
relativity, $\varkappa\rightarrow 0$, the cutoff vanishes:
\begin{equation}\label{11.2.5}
0\leq\omega^{2}\leq\infty
\end{equation}

\subsection{Product dynamics}

As long as the Hamiltonian is an (unbounded) selfadjoint
operator---due to the cutoff---it is possible to use product
dynamics [18-20]. It is this:
\begin{equation}\label{11.3.1}
\Psi(t_{2})=\hat{U}(t_{2},t_{1})\Psi(t_{1})
\end{equation}
\begin{equation}\label{11.3.2}
\hat{U}(t_{2},t_{1})=
T\exp\left\{-\mathrm{i}\int_{t_{1}}^{t_{2}}\hat{H}(t)dt\right\},\quad
\hat{H}^{\dag}(t)=\hat{H}(t)
\end{equation}
\begin{equation}\label{11.3.3}
\begin{array}{l}
T\exp\left\{-\mathrm{i}\int_{t_{1}}^{t_{2}}\hat{H}(t)dt\right\}=
\lim\limits_{N\rightarrow\infty}\mathrm{e}^{-\mathrm{i}\hat{H}(t_{2}-\Delta
t/N )\Delta t/N}\mathrm{e}^{-\mathrm{i}\hat{H}(t_{2}-2\Delta t/N
)\Delta t/N}\cdots \mathrm{e}^{-\mathrm{i}\hat{H}(t_{1})\Delta
t/N}\\ \Delta t=t_{2}-t_{1}>0
\end{array}
\end{equation}
Now
\begin{equation}\label{11.3.4}
\hat{U}^{\dag}(t_{2},t_{1})=T^{\dag}
\exp\left\{\mathrm{i}\int_{t_{1}}^{t_{2}}\hat{H}(t)dt\right\}=
\lim\limits_{N\rightarrow\infty}\mathrm{e}^{\mathrm{i}\hat{H}(t_{1}
)\Delta t/N}\cdots\mathrm{e}^{\mathrm{i}\hat{H}(t_{2}-\Delta t/N
)\Delta t/N}
\end{equation}
so that
\begin{equation}\label{11.3.5}
\hat{U}^{\dag}(t_{2},t_{1})\hat{U}(t_{2},t_{1})=
\hat{U}(t_{2},t_{1})\hat{U}^{\dag}(t_{2},t_{1})=I\;\;(\mathrm{unitarity}),\;\;
\hat{U}^{\dag}(t_{2},t_{1})=\hat{U}(t_{1},t_{2})
\end{equation}
This approach is better than the Dyson expansion [18-20] (in the
sense of convergence, not computability).

\section*{Acknowledgments}

I would like to thank Alex A. Lisyansky for support and Stefan V.
Mashkevich for helpful discussions.

\end{document}